\documentclass[12pt]{article}
\usepackage{amsfonts}
\usepackage{geometry}
\usepackage{graphicx}
\usepackage{amssymb}
\usepackage{epstopdf}
\usepackage{amsmath}
\usepackage{amssymb}
\usepackage{amsthm}

\begin{document}

\title{Realization of associative products in terms of Moyal and tomographic
symbols}
\author{A. Ibort$^a$, V.I. Man'ko$^b$, G. Marmo$^{c,d}$, A. Simoni$^{c,d}$,
C. Stornaiolo$^d$, \and and F. Ventriglia$^{c,d}$ \\
{\footnotesize \textit{$^a$Departamento de Matem\'{a}ticas, Universidad
Carlos III de Madrid, }}\\
{\footnotesize \textit{Av.da de la Universidad 30, 28911 Legan\'{e}s,
Madrid, Spain }}\\
{\footnotesize {(e-mail: \texttt{albertoi@math.uc3m.es})}}\\
{\footnotesize \textit{$^b$P.N.Lebedev Physical Institute, Leninskii
Prospect 53, Moscow 119991, Russia}}\\
{\footnotesize {(e-mail: \texttt{manko@na.infn.it})}}\\
\textsl{{\footnotesize {$^c$Dipartimento di Fisica dell' Universit%
\`{a} ``Federico II" ,}}}\\
\textsl{{\footnotesize {Complesso Universitario di Monte S. Angelo, via
Cintia, 80126 Naples, Italy}}}\\
\textsl{{\footnotesize {$^d$ Sezione INFN di Napoli,}}}\\
\textsl{{\footnotesize {Complesso Universitario di Monte S. Angelo, via
Cintia, 80126 Naples, Italy}}}\\
{\footnotesize {(e-mail: \texttt{marmo@na.infn.it, simoni@na.infn.it,
cosmo@na.infn.it, ventriglia@na.infn.it})}}}
\maketitle

\abstract{ The quantizer-dequantizer method allows to construct associative
products on any measure space. Here we consider an inverse problem: given an associative product is it possible
to realize it within the quantizer-dequantizer framework?
The answer is positive in finite dimensions and we give a few examples in infinite dimensions.}

\section{ Introduction}

The standard formulation of quantum mechanics associates pure states with
the state vector $|\psi \rangle $ \cite{Dirac} or wave function $\psi (x)$
\cite{Schr} while density states are described by density operators \cite%
{vonNeumann} \cite{landau} or density matrices. Observables are described by
Hermitian operators $\hat{A}=\hat{A}^{\dag }$ and their statistics, encoded
by their highest moments $\langle \hat{A}^{n}\rangle$ , is given by the
pairing of the state and the observable $\langle \hat{A}^{n}\rangle=\mathrm{%
Tr}[\hat{\rho}\hat{A}^{n}]$ \cite{vonNeumann,landau}. On the other hand in
classical statistical mechanics on phase-space the state is associated with
a probability density $f(q,p)$ and observables are functions $A(q,p)$ on
this space. The statistics encoded by the highest moments $\langle
A^{n}\rangle$ is described by the formulae of standard probability theory $%
\langle A^{n}\rangle=\int f(q,p)A^{n}(q,p)dqdp$. An attempt to find a
formulation of quantum mechanics which would proceed along the same rules of
classical statistics gave origin to the Weyl-Wigner formalism. Here operator
symbols \cite{Weyl27} are used as observables according to the Weyl map $%
A(q,p)\rightarrow \hat{A}$. The Wigner function \cite{Wigner32} $W(q,p)$
which is the Weyl symbol of the density operator $\hat{\rho}\rightarrow
W(q,p)$ has been introduced and used to describe quantum states. Since
operators form a noncommutative $C^{\star }$--algebra the Weyl map provides
a $C^{\star }$--algebra of functions in the phase space with a
noncommutative product--rule called \textquotedblright star-product". The
general construction of star--products was considered in many papers (see
\cite{Straton,Bayen:1977ha,Bayen:1977hb}). In connection with the recently
introduced tomographic picture of quantum mechanics (see \cite%
{ManciniPLA96,Ibort}) the star-product general scheme was discussed in
detail \cite{OlgaJPA} where the notion of quantizer and dequantizer
operators was introduced for arbitrary star product scheme. Mathematical
aspects in abstract form of the star--product in phase space were considered
in \cite{Fedosov:1994zz,Kontsevich:1997vb,Fairlie:2006uz}. In the
formulation of the star-product approach \cite{OlgaJPA} one considers
operators $\hat{D}(x)$ and $\hat{U}(x),$ acting on some Hilbert space $%
\mathcal{H}$, parametrized by points $x$ of a measure space $X$. $\hat{D}(x)$
and $\hat{U}(x)$ are called quantizer and dequantizer, respectively.

The bijective map $\hat{A}\leftrightarrow A(x)$ of operators onto their
symbols is given by taking the trace of the product of $\hat{A}$ with the
dequantizer. The reconstruction of operators from their symbols is given by
integration of the product $f_{A}(x)\hat{D}(x)$ over the measure space $X$.

The kernel of the noncommutative star--product of the symbols $A(x)$ and $%
B(x)$ of the operators $\hat{A}$ and ${\hat{B}}$ is determined by the
"structure constants" $K(x_1,x_2;x_3)=\mathrm{Tr}[\hat{D}(x_{1})\hat{D}%
(x_{2})\hat{U}(x_{3})]$. Thus, given the quantizer $\hat{D}(x)$ and the
dequantizer $\hat{U}(x)$ one obtains the star-product kernel. To the best of
our knowledge the inverse problem was not considered till now. Namely, given
a star-product kernel of functions on a manifold, is it possible to find a
pair of quantizers-dequantizers which allows to realize the star-product
kernel by the tracing formula? The aim of this paper is to obtain the
explicit equation for the quantizer if the star-product kernel is given. We
will show that such important example as Gr\"{o}newold kernel \cite{Gr}
provides the equation for finding the quantizer in the scheme of
Weyl--Moyal--Wigner symbols \cite{moyal49}. We also consider some other
known and unknown examples.

The paper is organized as follows. In the next Section \ref{sec2} we review
the construction of the star-product scheme following \cite{OlgaJPA} . In
Sect. \ref{sec3} the equation for the quantizer of the star-product scheme
is derived. In Sect. \ref{sec4} we study a known example of Moyal
star-product. In Sect. \ref{sec5} we apply the method to find the quantizer
for a discrete spin-system. In Sect. \ref{sec6} we consider the symplectic
tomographic map and the equation for the tomographic product kernel. In
Sect. \ref{sec7} we resume our results and perspectives.

\section{Quantizer--dequantizer pair and product}

\label{sec2}

Symbols of operators $\hat{A}$ and $\hat{B}$, determined by the dequantizer $%
\hat{U}(x)$ are given by
\begin{eqnarray}
f_{\hat{A}}(x) &=&\mathrm{Tr}[\hat{U}(x)\hat{A}]=A(x)  \label{1} \\
f_{\hat{B}}(x) &=&\mathrm{Tr}[\hat{U}(x)\hat{B}]=B(x)  \notag
\end{eqnarray}%
while the inverse are given by means of the quantizer $\hat{D}(x)$ as
\begin{eqnarray}
\hat{A} &=&\int f_{\hat{A}}(x)\hat{D}(x)dx  \label{2} \\
\hat{B} &=&\int f_{\hat{B}}(x)\hat{D}(x)dx  \notag
\end{eqnarray}%
provided that
\begin{equation}
\mathrm{Tr}[\hat{U}(x)\hat{D}(x^{\prime })]=\delta (x-x^{\prime }).
\label{3}
\end{equation}%
The star-product is defined by the kernel $K({x_{1}},{x_{2}},{x_{3}})$, i.e.
\begin{equation}
\left( f_{A}\ast f_{B}\right) (x_{3})=\int
K(x_{1},x_{2};x_{3})f_{A}(x_{1})f_{B}(x_{2})dx_{1}dx_{2}.  \label{4}
\end{equation}%
The kernel itself is determined by quantizer and dequantizer as \cite%
{OlgaJPA}
\begin{equation}
K(x_{1},x_{2};x_{3})=\mathrm{Tr}\left[ \hat{D}(x_{1})\hat{D}(x_{2})\hat{U}%
(x_{3})\right]  \label{5}
\end{equation}%
on the space $X.$ Out of this construction one obtains the bilinear, binary
associative product of functions $f_{A}(x)$, $f_{B}(x)$, $f_{C}(x)$ i.e.
\begin{equation}
\left( (f_{A}\ast f_{B})\ast f_{C}\right) (x)=\left( f_{A}\ast (f_{B}\ast
f_{C})\right) (x).  \label{6}
\end{equation}%
Property (\ref{6}) means that the kernel $K(x_{1},x_{2};x_{3})$ satisfies
the nonlinear (quadratic) equation
\begin{equation}
\int K({x_{1}},{x_{2}};{t})K({t},{x_{3}};{x_{4}})d{t}=\int K({x_{1}},{t};{%
x_{4}})K({x_{2}},{x_{3}};{t})d{t}.  \label{7}
\end{equation}

In \cite{OlgaPatrizia} the study of a dual star-product scheme was
considered. The pair $\hat{U}(x)$ and $\hat{D}(x)$ can be used to construct
the dual symbol of an operator $\hat{A}$ by exchanging the role of the
initial quantizer and dequantizer and considering the new dequantizer $\hat{U%
}^{d}(x)$ as
\begin{equation}
\hat{U}^{d}(x)=\hat{D}(x)  \label{D1}
\end{equation}%
and the new quantizer $\hat{D}^{d}(x)$ as
\begin{equation}
\hat{D}^{d}(x)=\hat{U}(x).  \label{D2}
\end{equation}%
These new operators satisfy the compatibility condition
\begin{equation}
\mathrm{Tr}[\hat{U}(x)\hat{D}(x^{\prime })]=\mathrm{Tr}[\hat{U}^{d}(x)\hat{D}%
^{d}(x^{\prime })]=\delta (x-x^{\prime }).  \label{D3}
\end{equation}%
In view of this, the dual symbol of an operator $\hat{A}$ reads
\begin{equation}
f_{A}^{d}(x)=\mathrm{Tr}[\hat{D}(x)\hat{A}]  \label{D4}
\end{equation}%
and the reconstruction formula provides an expression for the operator $\hat{%
A}$ in terms of its dual symbol
\begin{equation}
\hat{A}=\int f_{A}^{d}(x)\hat{U}(x)dx.  \label{D5}
\end{equation}%
The dual star-product kernel is given by the same formula (\ref{5}) with the
replacement $\hat{D}\leftrightarrow \hat{U}$, i.e.
\begin{equation}
K^{d}({x_{1}},{x_{2}};{x_{3}})=\mathrm{Tr}\left[ \hat{U}(x_{1})\hat{U}(x_{2})%
\hat{D}(x_{3})\right] .  \label{D6}
\end{equation}%
The meaning of the dual symbols and the dual star-product is based on the
possibility to express the mean value of a quantum observable $\hat{A}$ in
the form analogous to the formula of standard probability theory \cite%
{OlgaPatrizia,olga97}, i.e.
\begin{equation}
\langle \hat{A}\rangle =\mathrm{Tr}[\hat{\rho}\hat{A}]=\int \mathcal{W}(x)%
\mathcal{W}_{A}^{d}(x)dx.  \label{D7}
\end{equation}%
If $\mathcal{W}(x)$ is the symbol of the density operator $\hat{\rho}$ and
this symbol is such that it has the property of a fair probability
distribution like in the tomographic picture of quantum mechanics, the dual
symbol $\mathcal{W}_{A}^{d}(x)$ of an observable $\hat{A}$ plays the role of
the function identified with the observable in the star-product scheme under
consideration. Then the dual star-product kernel (\ref{D6}) provides a rule
of multiplication for the observables. The Weyl--Wigner--Moyal star--product
is self--dual since in this scheme $\hat{U}(x)=\lambda \hat{D}(x)$.

\section{Equations for the kernel and the quantizer}

\label{sec3}

We are now able to formulate the main problem of the present paper: Given
the associative product with kernel $K(x_{1},x_{2},x_{3})$, can we find the
pair $\hat{U}(x)$ and $\hat{D}(x)$ which provides the kernel by means of
equation (\ref{5})? We are searching for an equation for the pair $\hat{U}%
(x) $ and $\hat{D}(x)$. This equation can be obtained in the following way.
Let us first suppose, for a given kernel, that the unknown dequantizer $\hat{%
D}(x)$ exists. Then let us construct the operator
\begin{equation}
\hat{F}(x_{1},x_{2})=\int K(x_{1},x_{2};x_{3})\hat{D}(x_{3})dx_{3}.
\label{13}
\end{equation}%
The kernel can be interpreted as the symbol of the operator product $\hat{D}%
(x_{1})\hat{D}(x_{2}$) if one recalls equations (\ref{1}) and (\ref{5}).
Thus, due to the reconstruction formulae (\ref{2}), one has
\begin{equation}
\hat{F}(x_{1},x_{2})=\int K(x_{1},x_{2};x_{3})\hat{D}(x_{3})dx_{3}=\hat{D}%
(x_{1})\hat{D}(x_{2}).  \label{14}
\end{equation}%
In this formula the kernel is known and the quantizer $\hat{D}(x)$ is
unknown. It is just the equation which we are looking for.

Together with formula (\ref{3}), equation (\ref{14}) gives (in principle)
the pair quantizer $\hat{D}(x)$ and dequantizer $\hat{U}(x)$.

From our analysis for a given kernel of the dual star-product $%
K^{d}(x_{1},x_{2},x_{3})$ follows the equation for finding the operator $%
\hat{U}(x)$ which reads
\begin{equation}
\hat{U}(x_{1})\hat{U}(x_{2})=\int K^{d}(x_{1},x_{2};x_{3})\hat{U}%
(x_{3})dx_{3}.  \label{D8}
\end{equation}

In finite terms, we assume to have a vector space $V$, with a given basis $%
\{ v_{j}\}\ \ j=1,\dots n$, and structure constants for an associative
product $v_{j}\cdot v_{k}=\sum_l C_{jk}^{l}v_{l}$, our inverse problem
amounts to find matrices $D_{j}$ such that $D_{j}D_{k}=\sum_l
C_{jk}^{l}D_{l} $.

In the next section we illustrate this by using the Moyal-Gr\"onewold
product and the tomographic one.

\section{Solving the equation for the quantizer of known star-products}

\label{sec4}

Let us check now the validity of our Eq.(\ref{14}) for the Weyl product. We
put $\hbar =1.$The dequantizer for the Weyl symbol is the displaced parity
operator
\begin{equation}
\hat{U}(q,p)\equiv \hat{U}(z):=2\hat{\mathcal{D}}(z)\mathcal{\hat{P}}\hat{%
\mathcal{D}}^{\dagger }(z)=2\hat{\mathcal{D}}(2z)\mathcal{\hat{P}\ },\ z=%
\frac{q+ip}{\sqrt{2}},
\end{equation}%
where $\hat{\mathcal{D}}(z)=\exp \left[ z\hat{a}^{\dagger }-z^{\ast }\hat{a}%
\right] $\ is\ the usual displacement operator and $\mathcal{\hat{P}}=\exp %
\left[ i\pi \hat{a}^{\dagger }\hat{a}\right] $ is the parity operator.

The quantizer is
\begin{equation}
\hat{D}(q,p):=\frac{1}{2\pi }\hat{U}(q,p).  \label{B}
\end{equation}

Using $z_{k}=(q_{k}+ip_{k})/\sqrt{2},k=1,2,3,$\ the equation for the
quantizer determined by the Gr\"{o}newold kernel may be put in the form
\begin{eqnarray}
&&\exp \left[ 2(z_{1}^{\ast }z_{2}-z_{1}z_{2}^{\ast })\right] \int \exp %
\left[ 2z_{3}^{\ast }\left( z_{1}-z_{2}\right) -2z_{3}\left(
z_{1}-z_{2}\right) ^{\ast }\right] \hat{\mathcal{D}}(2z_{3})\frac{%
dq_{3}dp_{3}}{\pi }\mathcal{\hat{P}}  \notag \\
&=&\hat{\mathcal{D}}(2z_{1})\mathcal{\hat{P}}\hat{\mathcal{D}}(2z_{2})%
\mathcal{\hat{P}}.  \label{C}
\end{eqnarray}%
The integral above is the complex Fourier transform of the displacement
operator, which is known to be the displaced parity operator (see, e.g.,
eq.s (2.14) and (4.11) of \cite{Cahill-Glauber1969}). So, the l.h.s. of the
above equation\ becomes%
\begin{equation}
\exp \left[ 2(z_{1}^{\ast }z_{2}-z_{1}z_{2}^{\ast })\right] \hat{\mathcal{D}}%
(2\left[ z_{1}-z_{2}\right] )\mathcal{\hat{P}}^{2}=\hat{\mathcal{D}}(2z_{1})%
\hat{\mathcal{D}}(-2z_{2})\mathcal{\hat{P}}^{2}=\hat{\mathcal{D}}(2z_{1})%
\mathcal{\hat{P}}\hat{\mathcal{D}}(2z_{2})\mathcal{\hat{P}}.
\end{equation}%
This completes the check that eq.(\ref{14}) for the Moyal product kernel
provides the quantizer $\hat{D}(q,p)$ as solution.

\section{The case of discrete systems}

\label{sec5}

Now we check the validity of Eq. (\ref{14}) for one of the star-product
schemes with spin variables \cite{OlgaPatrizia}. Let us considere four Pauli
matrices
\begin{equation}
\sigma _{0}=%
\begin{pmatrix}
1 & 0 \\
0 & 1%
\end{pmatrix}%
,\,\ \sigma _{1}=%
\begin{pmatrix}
0 & 1 \\
1 & 0%
\end{pmatrix}%
,\,\sigma _{2}=%
\begin{pmatrix}
0 & -i \\
i & 0%
\end{pmatrix}%
,\,\sigma _{3}=%
\begin{pmatrix}
1 & 0 \\
0 & -1%
\end{pmatrix}%
.  \label{f1}
\end{equation}%
We recall their commutation relations
\begin{equation}
\lbrack \sigma _{0},\sigma _{j}]=0,\,\ \ \ \ \ \ [\sigma _{j},\sigma _{k}]=2
\sum _{m=1}^3 i\epsilon _{jkm}\sigma _{m}.  \label{f2}
\end{equation}%
The associative product reads
\begin{eqnarray}
\sigma _{j}\sigma _{0} &=&\sigma _{0}\sigma _{j}=\sigma _{j},  \label{f3} \\
\sigma _{j}\sigma _{k} &=&\delta _{jk}\sigma _{0}+\sum _{m=1}^3 i\epsilon
_{jkm}\sigma _{m}.  \label{f4}
\end{eqnarray}

Following \cite{OlgaPatrizia} in this section we use the pairing by the rule
$\left\langle \cdot ,\cdot \right\rangle =2\mathrm{Tr}\left[ \cdot \cdot %
\right] ,$ thus define the dequantizer $\hat{U}(x)$ \ for discrete label $%
x=\{0,1,2,3\}$ as the set of four matices $\hat{U}_{x}$
\begin{equation}
\left\{ \hat{U}_{0}=\frac{1}{2}\sigma _{0},\hat{U}_{1}=\frac{1}{2}\sigma
_{1},\hat{U}_{2}=\frac{1}{2}\sigma _{2},\hat{U}_{3}=\frac{1}{2}\sigma
_{3}\right\} .  \label{f5}
\end{equation}%
The quantizer we define as $\hat{D}_{x}=\hat{U}_{{x}}$. One has
\begin{equation}
\left\langle \hat{U}_{j}|\hat{D}_{k}\right\rangle =2\mathrm{Tr}[\hat{U}_{j}%
\hat{D}_{k}]=\delta _{jk}.  \label{f6}
\end{equation}%
The kernel of the star-product reads
\begin{equation}
K(j,k,m)=\frac{1}{4}\mathrm{Tr}[\sigma _{j}\sigma _{k}\sigma _{m}].
\label{f7}
\end{equation}%
This kernel can be represented in the form of four matrices, namely
\begin{eqnarray}
&&(K_{0})_{jk}=K(j,k,0),\ \ (K_{1})_{jk}=K(j,k,1),  \notag \\
&&(K_{2})_{jk}=K(j,k,2),\ \ (K_{3})_{jk}=K(j,k,3).  \label{k1}
\end{eqnarray}%
One can easily get these matrices in explicit form
\begin{eqnarray}
K_{0}=\frac{1}{2}%
\begin{pmatrix}
1 & 0 & 0 & 0 \\
0 & 1 & 0 & 0 \\
0 & 0 & 1 & 0 \\
0 & 0 & 0 & 1%
\end{pmatrix}%
, &&K_{1=}\frac{1}{2}%
\begin{pmatrix}
0 & 1 & 0 & 0 \\
1 & 0 & 0 & 0 \\
0 & 0 & 0 & i \\
0 & 0 & -i & 0%
\end{pmatrix}%
,  \label{k2bis} \\
K_{2}=\frac{1}{2}%
\begin{pmatrix}
0 & 0 & 1 & 0 \\
0 & 0 & 0 & -i \\
1 & 0 & 0 & 0 \\
0 & i & 0 & 0%
\end{pmatrix}%
, &&K_{3}=\frac{1}{2}%
\begin{pmatrix}
0 & 0 & 0 & 1 \\
0 & 0 & i & 0 \\
0 & -i & 0 & 0 \\
1 & 0 & 0 & 0%
\end{pmatrix}%
,  \notag
\end{eqnarray}

Of course, a solution for eq. (\ref{14}) is provided by the Pauli matrices,
indeed they satisfy the condition
\begin{equation}
\left( \frac{1}{2}\sigma _{j}\right) \left( \frac{1}{2}\sigma _{k}\right)
=\sum_{s=0}^{3}(K_{s})_{jk}\frac{\sigma _{s}}{2}.  \label{k3}
\end{equation}%
For example for $j=1,\,\,k=2$ from eq. (\ref{f4}) we get
\begin{equation}
\frac{1}{2}\sigma _{1}\cdot \frac{1}{2}\sigma _{2}=\frac{i\sigma _{3}}{4}.
\label{k4}
\end{equation}%
Another solution is provided by matrices $K_{0},K_{1},K_{2},K_{3}$ defined
in eq.(\ref{k2bis}). Thus we have provided two solutions for the equation (%
\ref{14}) for the quantizer matrices $\{D_j \}$. So, in conclusion, given
the structure constants, we search for the quantizer matrices $%
D_{1},D_{2},D_{3},D_{4}$ which would satisfy
\begin{equation}
D_{j}D_{k}=\sum_{l}C_{jk}^{l}D_{l}.
\end{equation}

To give an example where the structure constants are not primarily given by
a standard row--by--column product of matrices, we consider the following
product on $2\times 2-$matrices \cite{cargramar}
\begin{equation}
\begin{pmatrix}
a & b \\
c & d%
\end{pmatrix}%
\begin{pmatrix}
a^{\prime } & b^{\prime } \\
c^{\prime } & d^{\prime }%
\end{pmatrix}%
=%
\begin{pmatrix}
aa^{\prime } & ab^{\prime }+bd^{\prime } \\
ca^{\prime }+dc^{\prime } & dd^{\prime }%
\end{pmatrix}%
,  \label{A1}
\end{equation}%
one can check that this product is associative. Then, introducing the Weyl
basis of matrices (instead of Pauli matrices)
\begin{equation}
e_{1}=%
\begin{pmatrix}
1 & 0 \\
0 & 0%
\end{pmatrix}%
,\ \ e_{2}=%
\begin{pmatrix}
0 & 1 \\
0 & 0%
\end{pmatrix}%
,\ \ e_{3}=%
\begin{pmatrix}
0 & 0 \\
1 & 0%
\end{pmatrix}%
,\ \ e_{4}=%
\begin{pmatrix}
0 & 0 \\
0 & 1%
\end{pmatrix}%
,
\end{equation}%
we obtain the rule of multiplication
\begin{equation}
e_{j}\cdot e_{k}=\sum_{l=1}^{4}C_{jk}^{l}e_{l}  \label{|A2}
\end{equation}%
where, as one can see, only six components of the structure constants are
non zero, i.e.
\begin{equation}
C_{11}^{1}=C_{12}^{2}=C_{24}^{2}=C_{31}^{3}=C_{43}^{3}=C_{44}^{4}=1.
\label{A3}
\end{equation}%
Introducing functions on the four dimensional linear space in the form of a $%
4-$vectors
\begin{equation*}
\vec{f}=(f^{1},f^{2},f^{3},f^{4})
\end{equation*}%
where for any abstract vector $v=\sum_{j=1}^{4}v^{j}e_{j}$ the function $%
\vec{f}(v)=\sum_{j=1}^{4}v^{j}\vec{f}(e_{j})$ and $(\vec{f}%
(e_{j}))^{k}=\delta _{j}^{k}$ one has the star--product multiplication rule
for the functions $\vec{f}_{1}$ and $\vec{f}_{2}$
\begin{equation}
(\vec{f}_{1}\ast \vec{f}_{2})^{l}=%
\sum_{j,k=1}^{4}f^{j}_{1}C_{jk}^{l}f^{k}_{2}.  \label{A4}
\end{equation}%
The quantizer $4\times 4-$matrices are defined as $\left(D_{\gamma}\right)_{\beta}^{\alpha}=C_{\gamma\beta}^{\alpha}, (\alpha,\beta,\gamma=1,2,3,4),$ and read
\begin{eqnarray}
D_{1}=%
\begin{pmatrix}
1 & 0 & 0 & 0 \\
0 & 1 & 0 & 0 \\
0 & 0 & 0 & 0 \\
0 & 0 & 0 & 0%
\end{pmatrix}%
, &&D_{2}=%
\begin{pmatrix}
0 & 0 & 0 & 0 \\
0 & 0 & 0 & 1 \\
0 & 0 & 0 & 0 \\
0 & 0 & 0 & 0%
\end{pmatrix}%
,  \label{A2bis} \\
D_{3}=%
\begin{pmatrix}
0 & 0 & 0 & 0 \\
0 & 0 & 0 & 0 \\
1 & 0 & 0 & 0 \\
0 & 0 & 0 & 0%
\end{pmatrix}%
, &&D_{4}=%
\begin{pmatrix}
0 & 0 & 0 & 0 \\
0 & 0 & 0 & 0 \\
0 & 0 & 1 & 0 \\
0 & 0 & 0 & 1%
\end{pmatrix}%
.  \notag  \label{A5bis}
\end{eqnarray}%
The corresponding dequantizer matrices can be chosen solving the duality
condition
\begin{equation}
\mathrm{Tr}[D_{j}U_{k}]=\delta _{jk}.  \label{A6}
\end{equation}%
The quantizers and dequantizers must close on subalgebras of the general
linear group. Now one can see that by choosing
\begin{equation}
U_{1,4}=\frac{1}{2}D_{1,4}^{T}\ \ \ \ ,\ \ \ U_{2,3}=D_{2,3}^{T}  \label{A7}
\end{equation}%
one can check that while the $D$'s close on an algebra with structure
constants $C_{jk}^{l}$, the $U$'s close on an algebra with structure
constants $d_{jk}^{l}=\frac{1}{2}C_{kj}^{l}$. Then we found for the
considered exotic rule of multiplication of matrices the corresponding
star--product scheme with quantizers and dequantizers.

The considered example provides the star--product scheme with the kernel $%
C_{jk}^{l}$ given by the standard formula
\begin{equation}
C_{jk}^{l}=\mathrm{Tr}[D_{j}D_{k}U_{l}]
\end{equation}%
with quantizer (\ref{A2bis},\ref{A5bis}) and dequantizer given by (\ref{A7}).

A last example is given by considering the associative so--called $\kappa -$%
star--product \cite{OlgaJPA,MaMa97}, with the matrix multiplication rule $%
a\circ b=a\kappa b$. In case of $2\times 2-$matrices, by choosing a
Hermitian matrix $\kappa $, we may write
\begin{equation}
\kappa =\sum_{\alpha =0}^{3}s^{\alpha }\sigma _{\alpha }
\end{equation}%
where the components $s^{\alpha },\alpha =0,1,2,3,$ are real, and the $%
\sigma _{\alpha }$ are the previous Pauli matrices with the identity $\sigma
_{0}$.

The structure constants , with $\alpha =0,1,2,3,$ and $j,m,n=1,2,3$, are:%
\begin{eqnarray}
C_{00}^{\alpha } &=&s^{\alpha },C_{0j}^{\alpha }=\left( C_{j0}^{\alpha
}\right) ^{\ast }=\delta _{0}^{\alpha }s^{j}+\delta _{j}^{\alpha
}s^{0}+\delta _{m}^{\alpha }\sum_{n=1}^{3}is^{n}\epsilon _{njm}, \\
C_{jm}^{\alpha } &=&\delta _{0}^{\alpha }\left( s^{0}\delta
_{jm}+\sum_{n=1}^{3}is^{n}\epsilon _{nmj}\right) +\delta _{j}^{\alpha
}s^{m}+\delta _{m}^{\alpha }s^{j}+\delta _{n}^{\alpha }\left( i\
s^{0}\epsilon _{jmn}-\delta _{jm}s^{n}\right) .  \notag
\end{eqnarray}%
They give rise to the quantizer family:
\begin{eqnarray}
D_{0}=%
\begin{pmatrix}
s^{0} & s^{1} & s^{2} & s^{3} \\
s^{1} & s^{0} & is^{3} & -is^{2} \\
s^{2} & -is^{3} & s^{0} & is^{1} \\
s^{3} & is^{2} & -is^{1} & s^{0}%
\end{pmatrix}%
, &&D_{1}=%
\begin{pmatrix}
s^{1} & s^{0} & -is^{3} & is^{2} \\
s^{0} & s^{1} & -s^{2} & -s^{3} \\
-is^{3} & s^{2} & s^{1} & is^{0} \\
is^{2} & s^{3} & -is^{0} & s^{1}%
\end{pmatrix}%
, \\
D_{2}=%
\begin{pmatrix}
s^{2} & is^{3} & s^{0} & -is^{1} \\
is^{3} & s^{2} & s^{1} & -is^{0} \\
s^{0} & -s^{1} & s^{2} & -s^{3} \\
-is^{1} & is^{0} & s^{3} & s^{2}%
\end{pmatrix}%
, &&D_{3}=%
\begin{pmatrix}
s^{3} & -is^{2} & is^{1} & s^{0} \\
-is^{2} & s^{3} & is^{0} & s^{1} \\
is^{1} & -is^{0} & s^{3} & s^{2} \\
s^{0} & -s^{1} & -s^{2} & s^{3}%
\end{pmatrix}%
.  \notag
\end{eqnarray}
The dequantizers may be found by solving an equation like (\ref{A6}). We
conclude by observing that in the limit $\kappa \rightarrow \sigma_0$, i. e.
$s^0 \rightarrow 1, s^1,s^2,s^3 \rightarrow 0,$ the above matrices yield
just the matrices $K$'s of eq. (\ref{k2bis})

\section{Symplectic tomography}

\label{sec6} Now we prove that for the symplectic tomographic star-product
the quantizer and dequantizer satisfy the condition of compatibility for
homogeneous functions $f(X,\mu ,\nu )$. In fact the quantizer $\hat{D}(X,\mu
,\nu )$ and dequantizer $\hat{U}(X,\mu ,\nu ),$ say
\begin{equation}
\hat{D}(X,\mu ,\nu )=\frac{1}{2\pi }\exp i(X-\mu \hat{q}-\nu \hat{p}),\hat{U}%
(X,\mu ,\nu )=\delta (X-\mu \hat{q}-\nu \hat{p}),
\end{equation}%
give
\begin{equation}
\mathrm{Tr}\left[ \hat{U}(1)\hat{D}(2)\right] =\frac{1}{2\pi }\mathrm{Tr}%
\left[ \delta (X_{1}-\mu _{1}\hat{q}-\nu _{1}\hat{p})\exp i(X_{2}-\mu _{2}%
\hat{q}-\nu _{2}\hat{p})\right] ,  \label{F1}
\end{equation}%
which can be expressed by its action on homogeneous functions as
\begin{equation}
\frac{1}{(2\pi )^{2}}\int \mathrm{e}^{i(X^{\prime }-kX)}\delta (\mu ^{\prime
}-k\mu )\delta (\nu ^{\prime }-k\nu )f(X^{\prime },\mu ^{\prime },\nu
^{\prime })dX^{\prime }d\mu ^{\prime }d\nu ^{\prime }.  \label{F2}
\end{equation}%
Taking the Fourier transform with respect to the variable $X^{\prime }$ one
has the expression
\begin{equation}
\frac{1}{2\pi }\int \tilde{f}(-1,-k\mu ,-k\nu )e^{ikX}dk=f(X,\mu ,\nu ).
\label{F3}
\end{equation}%
We used the property of the Fourier transform of the homogeneous tomographic
symbol
\begin{equation}
\tilde{f}(k,\lambda \mu ,\lambda \nu )=\tilde{f}(k\lambda ,\mu ,\nu )
\label{F4}
\end{equation}

Thus we proved that the action of $\mathrm{Tr}[\hat{U}(x)\hat{D}(x^{\prime
})]$ onto the function $f(x),x=(X,\mu ,\nu ),$ which is homogeneous of
degree $-1,$ is equivalent to the integration of this function with the
Dirac delta-function $\delta (x-x^{\prime })$. However, the integration with
this function of non homogenous functions $F(X^{\prime },\mu ^{\prime },\nu
^{\prime })$ does not provide the same function $F(X,\mu ,\nu )$. This
property takes place also if we consider the solution of the equation for
the finding the quantizer of the tomographic star--product scheme. In fact
the relation
\begin{eqnarray}
&&\int \delta (\nu _{3}(\mu _{1}+\mu _{2})-\mu _{3}(\nu _{1}+\nu _{2}))\exp
\left[ -i\frac{\nu _{1}+\nu _{2}}{\nu _{3}}X_{3}\right] \hat{D}(X_{3},\mu
_{3},\nu _{3})dX_{3}d\mu _{3}d\nu _{3}  \notag \\
&&\times \frac{1}{4\pi ^{2}}\exp \left[ iX_{1}+iX_{2}+\frac{i}{2}(\nu
_{1}\mu _{2}-\nu _{2}\mu _{1})\right] =\hat{D}(X_{1},\mu _{1},\nu _{1})\hat{D%
}(X_{2},\mu _{2},\nu _{2})
\end{eqnarray}
holds true if one applies these operators to homogeneous functions.

\section{Conclusions}

\label{sec7}

We summarize the main results of our paper. Given the kernel of a
star-product which provides an associative product of functions on some
measure space $X$, is it possible to find a Hilbert space and a pair of
operator families, labeled by points of the space and called quantizer and
dequantizer, such that the kernel of the star--product is obtained by
tracing the product of two quantizers and one dequantizer? The answer which
we obtained is affirmative. The solution is provided by a nonlinear
equation, eq. (\ref{14}), for the quantizer operators for any given product
kernel.

We checked on the examples of the Moyal product, the tomographic product,
and the products defined on functions depending on discrete spin-variables
that there always exist solutions for the obtained equation for quantizers.
We conjecture that this situation takes place for arbitrary star-products of
functions both in finite and infinite spaces. In some sense we would
generalize the known result of Gelfand--Naimark--Segal which asserts that
one can always construct (GNS construction) a Hilbert space and the
operators which give a representation of a given $C^{\star }-$algebra. We
conjecture that all star--products on a measure space can be realized by
means of quantizers and dequantizers by the construction we have considered.
Our result may be considered also as an extension to associative algebras of
Ado's theorem available in the setting of Lie algebras, assuring that any
Lie product for abstract finite dimensional Lie algebra may be realized as
the commutator of matrices.

In a future paper we shall consider the method of contraction of associative
algebras by using a contraction procedure on quantizers and dequantizers.

Acknowledgment: V.I. Man'ko thanks the University "Federico II" of Naples
and INFN-Sezione di Napoli for their hospitality. G. Marmo would like to
acknowledge the support provided by the Santander/UCIIIM Chair of Excellence
programme 2011/2012.

\end{document}